# Limits on gas impermeability of graphene

P. Z. Sun[1,2], Q. Yang[1,2], W. J. Kuang[1], Y. V. Stebunov[1,2], W. Q. Xiong[3], J. Yu[3,4], R. R. Nair[2], M. I. Katsnelson[4], S. J. Yuan[3,4], I. V. Grigorieva[1], M. Lozada-Hidalgo[1], F. C. Wang[1,2,5] & A. K. Geim[1,2]

[1]School of Physics and Astronomy, University of Manchester, Manchester M13 9PL, UK
[2]National Graphene Institute, University of Manchester, Manchester M13 9PL, UK
[3]Key Laboratory of Artificial Micro- and Nano-structures of Ministry of Education and School of Physics and Technology, Wuhan University, Wuhan 430072, China
[4]Institute for Molecules and Materials, Radboud University, NL-6525 AJ Nijmegen, Netherlands
[5]Chinese Academy of Sciences Key Laboratory of Mechanical Behavior and Design of Materials, Department of Modern Mechanics, University of Science and Technology of China, Hefei, Anhui 230027, China

*Despite being only one-atom thick, defect-free graphene is considered to be completely impermeable to all gases and liquids[1-10]. This conclusion is based on theory[3-8] and supported by experiments[1,9,10] that could not detect gas permeation through micrometre-size membranes within a detection limit of $10^5$ to $10^6$ atoms per second. Here, using small monocrystalline containers tightly sealed with graphene, we show that defect-free graphene is impermeable with an accuracy of eight to nine orders of magnitude higher than in the previous experiments. We could discern permeation of just a few helium atoms per hour, and this detection limit is also valid for all other tested gases (neon, nitrogen, oxygen, argon, krypton and xenon), except for hydrogen. Hydrogen shows noticeable permeation, even though its molecule is larger than helium and should experience a higher energy barrier. The puzzling observation is attributed to a two-stage process that involves dissociation of molecular hydrogen at catalytically active graphene ripples, followed by adsorbed atoms flipping to the other side of the graphene sheet with a relatively low activation energy of about 1.0 electronvolt, a value close to that previously reported for proton transport[11,12]. Our work provides a key reference for the impermeability of two-dimensional materials and is important from a fundamental perspective and for their potential applications.*

From a theory standpoint, monolayer graphene imposes a very high energy barrier for penetration of atoms and molecules. Density-functional-theory calculations predict that the barrier $E$ is at least several eV[2-6], which should prohibit any gas permeation under ambient conditions. Indeed, one can estimate that at room temperature $T$ it would take longer than the lifetime of the universe to find an atom energetic enough to pierce a defect-free membrane of any realistic size. These expectations agree with experiments that reported no detectable gas permeation through mechanically-exfoliated graphene. The highest sensitivity was achieved using micrometer-size wells etched in oxidized silicon wafers, which were sealed with graphene[1,9,10]. In those measurements, a pressurized gas (e.g., helium) could permeate along the $SiO_2$ layer and gradually fill the microcontainers making so-called 'nanoballoons'. Their consecutive deflation in air could be monitored using atomic force microscopy (AFM), and it was shown that the leakage occurred only along the $SiO_2$ surface, within several minutes but independently of the number of graphene layers used for the sealing[1]. These studies allowed a conclusion that graphene membranes were impermeable to all gases, at least with the achieved accuracy of $10^5$–$10^6$ atoms s$^{-1}$. This was further corroborated by creating individual atomic-scale defects in graphene nanoballoons, which resulted in their relatively fast deflation/inflation and confirmed the high sensitivity of the method[9,10].

The devices used in our work were micrometer-size containers made from monocrystals of graphite or hexagonal boron nitride (hBN) using electron-beam lithography and dry etching (Fig. 1; *fig. S1*). The containers were sealed with graphene monolayer crystals obtained by mechanical exfoliation and transferred on top of the wells using van der Waals assembly ('Device fabrication' in Supplementary Information). In control experiments, bilayer graphene and monolayer $MoS_2$ were used for the sealing (see below). The wells were chosen to have an inner diameter $d$ of 0.5 or 1 µm whereas their depth $h$ was about 50 nm to minimize the containers' volume and, therefore, maximize the sensitivity with respect to the number of inflowing gas molecules. The depth could not be reduced further because van der Waals attraction of graphene to the inner walls caused its sagging[1,13,14], typically by a few tens of nanometers (Figs. 1c and d). The wells' ring-shaped top was typically 1-µm wide to provide a sufficiently large atomically-flat area so that no gas diffusion could occur along the resulting 'atomically-tight' sealing with its clean and sharp interface[15,16]. The monocrystalline walls of our microcontainers were also impermeable, as reported previously[17] and confirmed in the present work using wells with walls of different thicknesses. The rough surface outside the wells (due to etching) helped to pin the membranes preventing their slippage. The atomically-tight sealing is the principal difference with respect to the previous experimental setup[1,9,10] that used 'leaky' $SiO_2$. In our design, the only possible route for the gas ingress/escape is through the two-dimensional (2D) membrane.

The basic principle used for detection of molecular penetration through graphene membranes is similar to that introduced in ref. 1 and illustrated in Fig. 1a. The described microcontainers were placed inside a chosen gas atmosphere (e.g., He) and, if graphene were permeable to it, the partial pressures inside and outside should equalize so that the total pressure inside the containers (filled with the less permeating air) would increase with time, resulting in gradual lifting and eventual bulging of the membranes. Changes in the membrane position were monitored with AFM. All the containers were first checked for any possible damage to their sealing and the absence of atomic-scale defects[9,10] as described in 'Experimental procedures'. Only those containers that successfully passed those initial tests were investigated further. They were placed in He, initially for a few days. After that, the devices were taken out, measured by AFM within 10 minutes to detect minute changes in the membrane position $\delta$ (Fig. 1c) and quickly placed back for further exposure to He. To maximize our accuracy, AFM mapping was done using the PeakForce mode and, for repeated measurements, scans were taken in the same direction over the same area ('AFM measurements'). For the same reason, we minimized the stress imposed by different pressures inside and outside the containers by normally keeping the external pressure at $P_a$ = 1 bar and varying only the partial pressure $P$ of the tested gas (Supplementary Information). Furthermore, we avoided containers with $d$ larger than 1 µm because their scans appeared notably noisier ('AFM measurements'). Our experiments were limited to $T \leq 60\ °C$ because, after thermal cycling to higher $T$, graphene membranes were often destroyed, probably because of strain induced by thermal expansion/contraction (*fig. S2*).

Under exposure to helium, no changes in $\delta$ could be detected as detailed in Figs. 1d-f and *fig. S3*. The figures basically show that membrane positions did not change regardless of how long they were exposed to He. For example, Fig. 1e plots our results for more than a dozen of containers over an observation period of one month. None of the devices showed any discernable changes ($\Delta\delta$) in the membranes' original positions, beyond small random fluctuations that did not exceed 0.5 nm in amplitude. Their statistical analysis yielded the standard deviation (SD) of ~1 Å (Fig. 1e; 'AFM measurements'). In control experiments, we carried out the same measurements in air and found fluctuations of similar amplitude (*fig. S3*). For higher applied $P$, the fluctuations were slightly stronger, presumably because the extra pressure caused membranes' creeping (*fig. S3*).

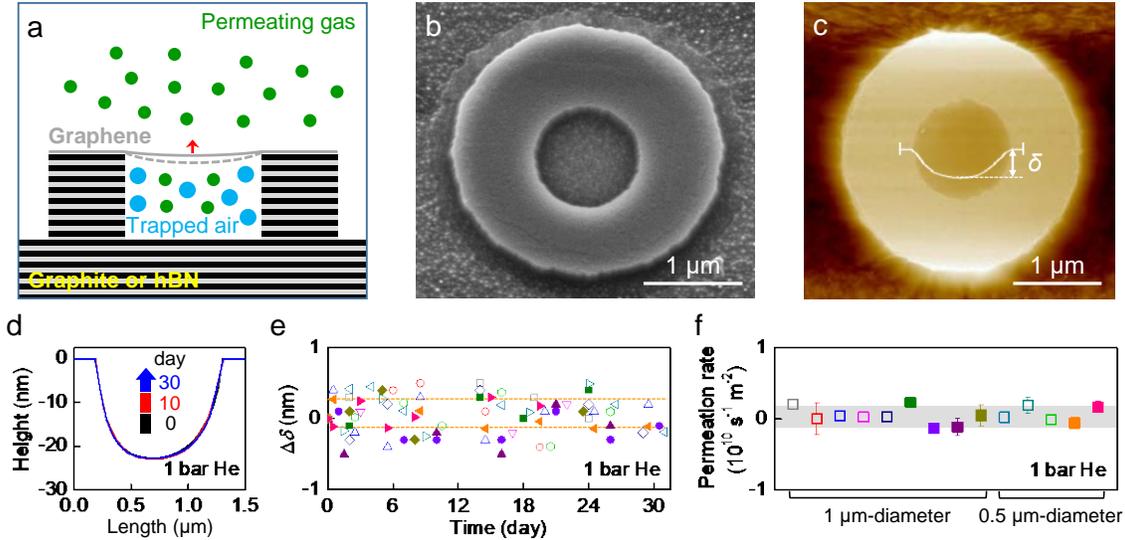

**Figure 1| Graphene's impermeability to helium. a,** Schematic of our experimental setup. **b,** Electron micrograph of one of the studied containers. The image was taken at a tilt angle of 20° for better view. The graphene membrane is seen to stretch over the outer wall and attach to the dry-etched surface outside. **c,** AFM image of a similar device. White curve: Profile of the suspended graphene along the well's diameter. The vertical bars indicate the width (~150 nm) over which such profiles were averaged. **d,** Examples of AFM profiles for the same container after storing it in He for days (color coded). **e,** Changes in the maximum deflection point for 14 containers placed in He over a one-month period. Different symbols denote containers made from graphite (empty symbols) and hBN (solid). The orange lines indicate the experimental scatter (full range of observed $\Delta\delta$) for one of the devices represented by the same color. **f,** Permeation rates $\Gamma$ evaluated from the data in (e); same symbol coding. Error bars: SD by fitting $\delta$ with the linear time dependence. The grey area indicates the overall SD using the data for all the devices.

For small changes $\Delta\delta$ in the membrane position, the number of atoms or molecules $\Delta N$ penetrating through the area $S$ is given by[9,18]

$$\Delta N = c \frac{P_a}{k_B T} S \Delta\delta \quad (1)$$

where $k_B T$ is the thermal energy, and $c \approx 0.5$ is the coefficient that accounts for the membrane's curved profile ('Evaluation of permeation rates and their accuracy'). The above accuracy of ~1 Å over a month translates into no more than a few atoms entering a microcontainer per hour. This accuracy is more than 8 orders of magnitude higher than that achieved in the earlier experiments reporting graphene's impermeability[1,9,10], which were in turn a few orders of magnitude more sensitive than the detection limit of modern helium leak detectors. In terms of the areal permeation rates $\Gamma = (d\Delta N/dt)/S$, our experiments yield the upper bound of ~$10^9$ s$^{-1}$ m$^{-2}$ for possible helium transparency of defect-free graphene. To put this into perspective, monolayer graphene is less permeable than one-kilometer-thick quartz glass. Furthermore, the found limit allows a lower-bound estimate for the energy barrier $E$ that graphene presents for helium atoms. Using the expression ('Energy barriers' in Supplementary Information)

$$\Gamma = \frac{P}{\sqrt{2\pi m k_B T}} \exp\left(-\frac{E}{k_B T}\right) \quad (2)$$

we obtain $E \geq 1.2$ eV, where the pre-exponential factor describes the incident rate of He atoms, and $m$ is their mass. This estimate is consistent with the barriers found theoretically[2-4]. Keeping in mind that it is hardly possible to improve the accuracy for $\Delta\delta$ beyond 1 Å whereas the fluctuations increase for larger $d$ and that the observations longer than a few months and at considerably higher *T* are impractical, our results probably present the limit for sensitivity of the nanoballoon method.

Helium with its small weakly-interacting atoms is recognized as the most permeating of all gases. Nonetheless, we tested several other gases (namely, Ne, Ar, $O_2$, Kr, $N_2$ and Xe) and, as expected, found no discernable permeation. This places practically the same limit on their *E*. Unexpectedly, monolayer graphene exhibited noticeable transparency with respect to molecular hydrogen $H_2$. We first illustrate this observation qualitatively, by showing in Fig. 2a one of our microcontainers before and after its exposure to hydrogen at 50°C for three days. The membrane clearly bulged up, although the same container passed our impermeability tests with respect to both He and Ar at the same *T*. This observation is striking because even atomic hydrogen, with its diameter smaller than that of helium, is predicted to experience *E* of 2.6–4.6 eV for monolayer graphene[4-6], leaving aside the fact that dissociation of $H_2$ requires ~4.5 eV which makes the concentration of atomic hydrogen negligible. Molecular hydrogen is expected[3] to have even higher *E* >10 eV. For so high barriers, hydrogen permeation is completely forbidden and, as per eq. (2), it should take billions of years for a single hydrogen atom to get inside the container. In another control experiment, we used microcontainers sealed with bilayer graphene and monolayer $MoS_2$. They exhibited no detectable permeation under long exposure to $H_2$ at 50°C (*fig. S4*).

To quantify the observed hydrogen permeation, we measured changes in $\delta$ as a function of time for many devices at room *T* (295 ±2 K). They exhibited approximately the same inflation rates within scatter of about ±15% as indicated by the dashed lines in Fig. 2b, which yields $\Gamma \approx 2 \times 10^{10}$ s$^{-1}$ m$^{-2}$. Note that such a minute gas influx is far beyond the detection limit for microcontainers with $SiO_2$ sealing[1,9,10]. Furthermore, working in the regime of small linear-in-time $\Delta\delta$ (no bulging as in Fig. 2a), we measured hydrogen permeation at different *T*. The temperature dependences followed the Arrhenius law, $\Gamma \propto \exp(-E/k_BT)$ yielding an activation barrier of 1.0 ±0.1 eV (Fig. 2c). This relatively small *E* strongly disagrees with the theory expectations and, more importantly, with the fact that smaller helium atoms did not penetrate through the same membranes. Trying to understand the origin of the unexpected behavior, we performed two additional sets of experiments. First, we quantified the hydrogen permeation rates at different pressures *P* and found $\Gamma \propto P^{1/2}$ (*fig. S5*). The square-root dependence is characteristic of processes involving an equilibrium between adsorbed and desorbed gas particles[19], in contrast to the linear dependence of eq. (2) valid for weakly-interacting atoms ('Energy barriers'). Second, we measured permeation for hydrogen's isotope deuterium. Within our detection limit, no permeation could be discerned, which puts a limit of $\Gamma \leq 10^9$ s$^{-1}$ m$^{-2}$ on the deuterium influx ('Isotope effect' and *fig. S8*).

To understand the reason for hydrogen's exclusivity among the other gases, let us recall the following facts. Locally-curved and strained graphene surfaces are known experimentally to be chemically reactive[20,21] and are expected to lower the energy required for dissociation of molecular hydrogen[22,23]. For a local protrusion (ripple) with $t/D \approx 5\%$ (where *t* is its height and *D* the lateral size), the dissociated state with two hydrogen adatoms becomes energetically more favorable[22,23], whereas the energy barrier required to reach this state is also reduced to ~1 eV ('Ab initio simulations of graphene's catalytic activity'). This catalytic activity of graphene is relevant because suspended membranes exhibit extensive nanoscale rippling[24-26] with typical $t/D$ that easily reaches ~5% for both static[24,25] and dynamic[26] ripples (*fig. S7*). Another fact is that monolayer graphene is known to be highly permeable to protons, exhibiting an activation energy of 1.0 ±0.05 eV, whereas bilayer graphene and monolayer $MoS_2$ exhibit no detectable proton permeation[11,12]. This is relevant because a hydrogen atom absorbed on graphene shares its electron with the conducting surface and is indistinguishable from an adsorbed proton.

Furthermore, it is also known that deuterons, nuclei of deuterium atoms, experience a higher barrier than protons, which drastically slows their permeation through monolayer graphene[12] ('Isotope effect').

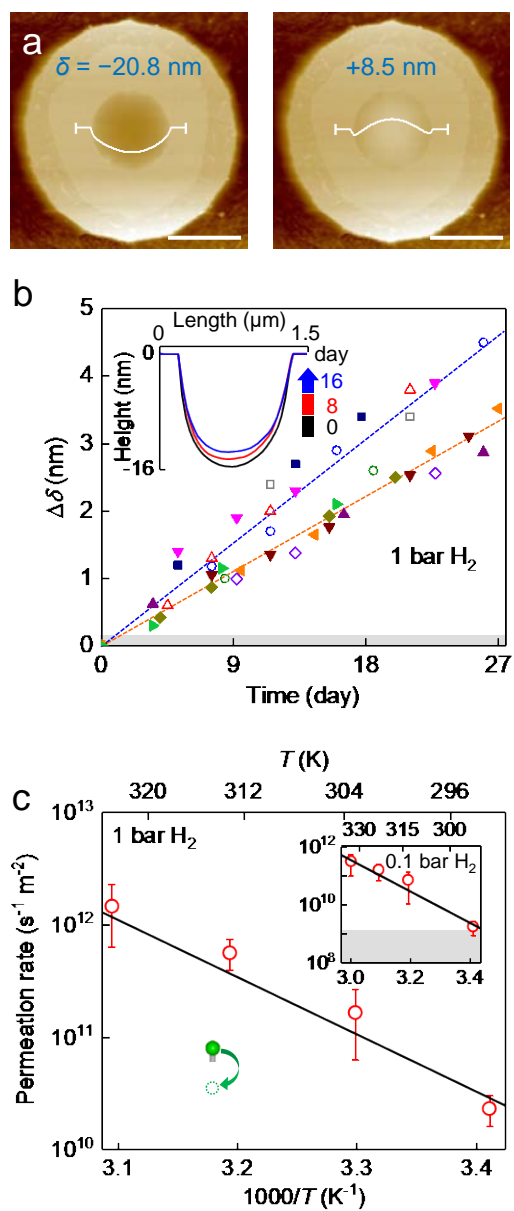

**Figure 2| Hydrogen permeation through defect-free graphene. a**, AFM micrographs of the same microcontainer before (left panel) and after (right) storing it for 3 days in $H_2$ at 1 bar. To speed up permeation, the gas was heated to 50 °C. White curves: Height profiles along the well's diameter. Tens of microcontainers were tested, showing the same effect. Scale bar, 1 μm. The somewhat darker outside region of the well's top appears because it is not atomically flat but slightly tapered (our lithography masks often thinned towards the outside perimeter, allowing some plasma etching of the rim region). **b**, Time evolution of $\Delta\delta$ for 12 different devices in 1 bar $H_2$ at $T = 295 \pm 2$ K. The empty and solid symbols denote graphite and hBN wells, respectively. Blue and orange dashed lines: Best linear fits for two of the devices (color coded) to indicate experimental scatter. Inset: Representative changes in the AFM profiles with time. **c**, Hydrogen permeation rates at different T. Symbols: Experimental data. Solid curve: Best fit to the activation behavior yields $E = 1.0 \pm 0.1$ eV. Top inset: Same for P = 0.1 bar. Error bars: SDs using 6 or more devices for each T. Bottom inset: Illustration of the flipping process in the suggested mechanism of hydrogen permeation. The grey areas in (b, c) indicate our detection limit.

Based on the above facts, we propose the following scenario for the observed hydrogen permeation. First, $H_2$ is chemisorbed (adsorbed and dissociated) on graphene ripples, which results in $sp^3$-bonded adatoms as illustrated in *fig. S6*. These adatoms then flip to the other side of graphene in the same 1.0-eV transfer process that was previously reported for proton transport[11,12] (inset of Fig. 2c). The flipped adatoms subsequently desorb from the concave surface. This scenario is fully consistent with all the experimental evidence and, also, explains why the observed permeation is limited to hydrogen and monolayer graphene: Among the tested 2D crystals, only the latter is transparent to protons. Neither bilayer graphene nor monolayer $MoS_2$ allows protons through[11], whereas even monolayer graphene presents a notably higher barrier for heavier deuterons[12] (see 'Isotope effect').

Although our experiments cannot distinguish directly whether it is chemisorption or flipping that limits the hydrogen permeation, the close match of the measured *E* with that reported in ref.[12] hints that the flipping is likely to be the rate limiting process. This is supported by the observed isotope effect. Indeed, our density-functional-theory calculations could not find any influence of zero-point oscillations on hydrogen's dissociation ('Isotope effect'). On the other hand, the flipping is expected to exhibit an isotope shift because zero-point oscillations decrease the bonding energy of hydrogen adatoms, which characterizes the initial state in the transfer process[12]. This shift should make deuterium permeation sufficiently slow to become undetectable in our experiments ('Isotope effect'). The dependence $\Gamma \propto P^{1/2}$ suggesting a finite coverage of graphene with hydrogen is also consistent with the flipping being the limiting step. Indeed, it is easier for lighter adatoms to desorb from graphene (because of stronger zero-point oscillations), which should result in higher coverage of the graphene surface with deuterium. Accordingly, if chemisorption were the limiting step, higher permeation rates would be expected for $D_2$ rather than $H_2$, contrary to our observations.

To conclude, defect-free graphene should be completely impermeable to all atomic species. However, ripples, wrinkles and other defects inducing a local curvature allow a non-negligible throughput of hydrogen. If necessary, the latter permeation can be blocked by using bilayer graphene or other 2D materials such as $MoS_2$. Our results have implications for many observations in the literature. For example, ripples probably play an important role in lowering barriers for proton transport through 2D membranes[11,12], a distinct possibility not considered so far in calculations[7,27-29]. Similarly, the observations may shed light on intercalation of graphene-on-SiC by molecular hydrogen that was argued to permeate through defect-free graphene[30,31]. The discussed processes may also be critical for the interaction of graphene with water and hydrocarbons and, more generally, emphasize high catalytic activity of non-flat graphene, in stark contrast to the extreme chemical inertness of its bulk counterpart, graphite.

1. Bunch, J. S. *et al. Nano Lett.* **8**, 2458–2462 (2008).
2. Berry, V. *Carbon* **62**, 1–10 (2013).
3. Leenaerts, O., Partoens, B. & Peeters, F. M. *Appl. Phys. Lett.* **93**, 193107 (2008).
4. Tsetseris, L. & Pantelides, S. T. *Carbon* **67**, 58–63 (2014).
5. Miao, M., Nardelli, M. B., Wang, Q. & Liu, Y. *Phys. Chem. Chem. Phys.* **15**, 16132–16137 (2013).
6. Seel, M. & Pandey, R. *2D Mater.* **3**, 025004 (2016).
7. Feng, Y. *et al. J. Phys. Chem. Lett.* **8**, 6009–6014 (2017).
8. Wang, W. L. & Kaxiras, E. *New J. Phys.* **12**, 125012 (2010).
9. Koenig, S. P., Wang, L., Pellegrino, J. & Bunch, J. S. *Nat. Nanotechnol.* **7**, 728–732 (2012).

10. Wang, L. *et al. Nat. Nanotechnol.* **10**, 785–790 (2015).
11. Hu, S. *et al. Nature* **516**, 227–230 (2014)
12. Lozada-Hidalgo, M. *et al. Science* **351**, 68–70 (2016).
13. Bunch, J. S. *et al. Science* **315**, 490–493 (2007).
14. Radha, B. *et al. Nature* **538**, 222–225 (2016).
15. Haigh, S. J. *et al. Nat. Mater.* **11**, 764–767 (2012).
16. Kelly, D. J. *et al. Nano Lett.* **18**, 1168–1174 (2018).
17. Hu, S. *et al. Nat. Nanotechnol.* **13**, 468–472 (2018).
18. Koenig, S. P., Boddeti, N. G., Dunn, M. L. & Bunch, J. S. *Nat. Nanotechnol.* **6**, 543–546 (2011).
19. Deveau, N. D., Ma, Y. H. & Datta, R. *J. Memb. Sci.* **437**. 298–311 (2013).
20. Wu, Q. *et al. Chem. Commun.* **49**, 677–679 (2013).
21. Bissett, M. A., Konabe, S., Okada, S., Tsuji, M. & Ago, H. *ACS Nano* **7**, 10335–10343 (2013).
22. Boukhvalov, D. W. & Katsnelson, M. I. *J. Phys. Chem. C* **113**, 14176–14178 (2009).
23. McKay, H., Wales, D. J., Jenkins, S. J., Verges, J. A. & de Andres, P. L. *Phys. Rev. B* **81**, 075425 (2010).
24. Meyer, J. C. *et al. Solid State Commun.* **143**, 101–109 (2007).
25. Geringer, V. *et al. Phys. Rev. Lett.* **102**, 076102 (2009).
26. Fasolino, A., Los, J. H. & Katsnelson, M. I. *Nat. Mater.* **6,** 858–861 (2007).
27. Kroes, J. M. H., Fasolino, A. & Katsnelson, M. I. *Phys. Chem. Chem. Phys.* **19**, 5813–5817 (2017).
28. Poltavsky, I., Zheng, L., Mortazavi, M. & Tkatchenko, A. *J. Chem. Phys.* **148**, 204707 (2018).
29. Mazzuca, J. W. & Haut, N. K. *J. Chem. Phys.* **148**, 224301 (2018).
30. Riedl, C., Coletti, C., Iwasaki, T., Zakharov, A. A. & Starke, U. *Phys. Rev. Lett.* **103**, 246804 (2009).
31. Kunc, J., Rejhon, M. & Hlídek, P. *AIP Adv.* **8**, 045015 (2018).
32. Geim, A. K. & Grigorieva, I. V. *Nature* **499**, 419–425 (2013).
33. Wang, L. *et al*. *Science* **342**, 614–617 (2013).
34. Park, H. G. & Jung, Y. *Chem. Soc. Rev.* **43**, 565-576 (2014).
35. Whittaker, J. D., Minot, E. D., Tanenbaum, D. M., McEuen, P. L. & Davis, R. C. *Nano Lett.* **6**, 953–957 (2006).
36. Hencky, H. *Z. fur Mathematik und Physik* **63**, 311–317 (1915).
37. Wang, G. *et al. Phys. Rev. Lett.* **119**, 036101 (2017).
38. Landau, L. D. & Lifshitz, E. M. Course of theoretical physics. Volume 5 (1980).
39. Kresse, G. & Furthmuller, J. *Phys. Rev. B* **54,** 11169–11186 (1996).
40. Perdew, J. P., Burke, K. & Ernzerhof, M. M. *Phys. Rev. Lett.* **77**, 3865–3868 (1996).
41. Monkhorst, H. J. & Pack, J. D. *Phys. Rev. B* **13**, 5188–5192 (1976).
42. Grimme, S. *J. Comput. Chem.* **27**, 1787–1799 (2006).
43. Kerber, T., Sierka, M. & Sauer, J. *J. Comput. Chem.* **29**, 2088–2097 (2008).
44. Sheppard, D., Xiao, P., Chemelewski, W., Johnson, D. D. & Henkelman, G. *J. Chem. Phys.* **136**, 074103 (2012).
45. Herzberg, G. & Monfils, A. *J. Mol. Spectrosc.* **5**, 482–498 (1961).
46. Plimpton, S. *J. Comput. Phys.* **117**, 1–19 (1995).
47. Hornekaer, L. *et al. Phys. Rev. Lett.* **97**, 186102 (2006).
48. Paris, A. *et al. Adv. Funct. Mater.* **23**, 1628–1635 (2013).
49. Bukola, S., & Creager, S. E. *Electrochimica Acta* **296**, 1–7 (2019).

## Supplementary Information

**Device fabrication.** Our devices were fabricated as shown schematically in *fig. S1a*. A monocrystal of either graphite (*NGS Naturgraphit*) or hBN (*HQ Graphene*) with a thickness of at least 150 nm was first mechanically exfoliated onto an oxidized silicon wafer that was freshly cleaned in oxygen plasma. The quality of the crystal's top surface was carefully checked for the presence of atomic terraces using dark-field and differential-interference-contrast microscopy. These modes allow detection of crystal edges and tears, even for a monolayer thickness. Using e-beam lithography, a set of ring-shaped polymer masks with an inner diameter $d$ of 0.5 or 1 μm was patterned on atomically-flat parts of the surface (without terraces). Reactive-ion etching was then used to remove ~50 nm of the exposed area to form micrometer-diameter wells (*fig. S1b*). After the lithography mask was dissolved, we annealed the structures in a hydrogen-argon (1:10) atmosphere at 400°C for 6 hours. Then a relatively large crystal of monolayer graphene (also obtained by mechanical exfoliation) was transferred in air on top of the wells using the transfer procedures standard for assembly of van der Waals heterostructures[32,33]. *Fig. S1b* shows an optical image of an array of graphene-sealed hBN wells. Closer views of such microcontainers are provided in Fig. 1 and *fig. S2a*, where one can clearly see graphene membranes draping over the wells. For comparison, *fig. S2b* shows a broken graphene membrane after our unsuccessful attempt to test it at 80°C. Because inner walls of the containers are not perfectly round (see, e.g., Figs. 1b and c), graphene membranes sag inside in a slightly asymmetric manner as noticeable in some AFM profiles (e.g., Fig. 1d).

**Experimental procedures.** After the fabrication, microcontainers were first checked with AFM for possible tears, wrinkles or other defects. Only devices with seemingly perfect sealing were used for further investigation. Those were tested further by placing them in a 3-bar argon atmosphere overnight. Occasionally, we found inflated containers that deflated quickly in air. In principle, this could be due to defects[9,10,34] but in most cases we could trace the leakage to poor sealing of the microcontainers: Either wells' top surface was slightly damaged by dry etching so that the rough streaks connected the inner and outer rim edges or small wrinkles were present, as retrospectively revealed by dedicated AFM analysis and scanning electron microscopy. The devices that passed the above tests were placed in a tested gas atmosphere and their possible inflation was carefully monitored as described in the main text.

For gas tests, microcontainers were placed inside a small stainless-steel chamber. It was evacuated to ~$10^{-3}$ mbar and then an investigated gas was introduced inside. For studies of temperature dependences, the whole vacuum chamber was placed inside an oven with controllable $T$.

**AFM measurements.** To monitor changes in position of graphene membranes, we employed the PeakForce mapping mode (*Dimension FastScan* from *Bruker*). The use of this AFM mode was essential to achieve the highest possible accuracy in our measurements of the membrane position. The PeakForce mode minimizes tip-to-sample interactions by employing an imaging force that can be as small as ~1 nN and has little effect on suspended graphene. For comparison, if we tried the contact-mode AFM imaging, graphene membranes were found to sag down after each scan by as much as several nanometers, which was obviously unacceptable for our purposes. The PeakForce mode also allows straightforward analysis of the obtained scans, as compared to the tapping AFM mode where a non-negligible pressure induced by the tip requires rather involved deconvolution of AFM images[1,35]. Furthermore, to maximize reproducibility between consecutive PeakForce scans, they were taken always along the same direction $x$ across the center of suspended graphene membranes and averaged over a finite width of ~150 nm (vertical bars in the shown AFM images). This approach also allowed us to increase accuracy by avoiding changes in AFM profiles caused by the slight asymmetry in membranes' sagging, which was pointed out in 'Device fabrication'. The asymmetry remains constant during measurements for a given device and does not affect our results. The averaged deflection profile $\delta(x)$ allowed us to detect minor changes $\Delta\delta$ in membrane's lowest position at the well's center (denoted previously as $\delta$) such that $\delta = \delta(0) + \Delta\delta$ where $\delta(0)$ is the initial position at the well's center.

*Figs. S3a and S3b* show the accuracy and reproducibility of our measurements of $\delta(x)$ and $\Delta\delta$. For this data set, two microcontainers were measured in air as described above, and 10 AFM scans were taken at one-hour intervals. Between each scan the devices were taken out of the AFM setup and then placed back to mimic the real measurement procedures. The figures show that the profiles $\delta(x)$ were very stable, and the resulting $\Delta\delta$ did not exceed ~0.3 nm. The statistical uncertainty (SD) for this set of AFM measurements was ~0.16 nm. To check for longer-term stability, 12 microcontainers with $d$ = 0.5 and 1 μm and different sagging (maximum depth $\delta$ varied between 5 and 25 nm) were kept in air for more than 20 days. Their height profiles were captured at regularly, a few days apart. The devices also exhibited excellent stability such that $\Delta\delta$ did not exceed 0.3 nm (*fig. S3c*), in good agreement with the short-term results in *fig. S3a*.

After exposing microcontainers to higher He pressures, we again did not observe any discernable changes in $\delta(x)$ but random fluctuations in $\Delta\delta$ somewhat increased (*figs. S3e and S3f*), presumably because of additional stresses induced by pressure. Note that, trying to improve our measurement accuracy further, we also made and tested microcontainers with $d \geq 2$ μm. However, their stability was significantly worse, with $\Delta\delta$ exceeding 1 nm, probably due to increasing instabilities caused by tip-membrane interactions. Data from such wells were not used in the reported analysis.

**Evaluation of permeation rates and their accuracy.** Eq. (1) of the main text can be deduced from the expression derived in ref. 9 as follows. The pressure $P$ inside the container includes two components: One is the initial atmospheric pressure of the trapped air ($P_a$) and the other is $\Delta P$, the pressure change induced by molecular permeation. For small changes in the membrane position, $\Delta P$ can be estimated from the Hencky solution[36] as

$$\Delta P = K(\nu) \cdot E \cdot L \cdot \Delta\delta^3 / a^4 \tag{S1}$$

where $E$ is Young's modulus, $L$ is the membrane thickness, $a$ is the radius of the container and $K(\nu)$ is the coefficient that depends on Poisson's ratio $\nu$. For graphene[18,37], $E$ = 1 TPa, $\nu$ = 0.16 and $K(\nu)$ = 3.09. The gas volume inside the container is given by $V = V_0 + \Delta V = S \cdot h + c \cdot S \cdot \delta$, where $S$ and $h$ are the container's area and depth, respectively, and $c \approx 0.5$ is the numerical coefficient accounting for the curved shape of the membrane. Putting the above expressions for $P$ and $V$ into the ideal gas law $PV = (N_0 + \Delta N) k_B T$ ($N_0$ is the initial number of air molecules), we find

$$\Delta N = (P_a \Delta V + V_0 \Delta P + \Delta P \Delta V)/k_B T \tag{S2}$$

Substituting the expressions for $\Delta V$ and $\Delta P$, eq. (S2) can be written as[9]

$$\Delta N = \frac{S}{k_B T}\left[cP_a \Delta\delta + \left(h + c\delta(0)\right)\frac{K(\nu)EL}{a^4}(\Delta\delta)^3 + c\frac{K(\nu)EL}{a^4}(\Delta\delta)^4\right] \tag{S3}$$

For the known constants and noticing that the largest deflection $\Delta\delta$ used in our quantitative analysis was only ~4 nm, we find that the second and third terms of eq. (S3) should not exceed 20% of the linear term's value. Therefore, for the purpose of our analysis, eq. (S3) can be simplified as eq. (1) used in the main text. This also agrees with the fact that $\Delta\delta$ evolved linearly in time, within our experimental scatter (see Fig. 2b). If contributions of the nonlinear terms were considerable, $\Delta\delta$ should start evolving nonlinearly as a function of $\Delta N$ and, hence, time.

**Energy barriers.** Helium permeation through the barrier presented by a graphene membrane can be estimated using

$$\Gamma = A\exp(-\frac{E}{k_B T}) \qquad (S4)$$

where $E$ is the energy barrier for incident atoms and $A$ is their attempt rate (that is, the number of atoms striking a unit area per second). Weakly-interacting He atoms cannot get adsorbed onto graphene and, therefore, the attempt rate is given by[38]

$$A = \frac{1}{4}\frac{N}{V}\langle v \rangle = \frac{P}{4\,k_B T}\langle v \rangle \qquad (S5)$$

where $\langle v \rangle = \sqrt{8k_B T/\pi m}$ is the mean speed of helium atoms and $m$ is their atomic weight. Combining eqs. (S4-S5), we obtain eq. (2) of the main text. On the other hand, if gas atoms or molecules become adsorbed on a graphene surface, like in the case of hydrogen, eq. (S5) is no longer applicable, and the attempt rate depends on an equilibrium density of adsorbed species. Under the latter circumstances, a different pressure dependence $A \propto \sqrt{P}$ is expected for bipartite-gas dissociation[19], in agreement with our results for hydrogen in *fig. S5*.

**Ab initio simulations of graphene's catalytic activity.** Energy barriers for dissociation of molecular hydrogen $H_2$ on flat and rippled graphene were calculated from first principles using the density functional theory (DFT), as implemented in Vienna *ab initio* package (VASP)[39]. The generalized gradient approximation (GGA)[40] and projected augmented wave (PAW) were adopted to describe the exchange correlation potential and ion-electron interactions. The kinetic energy cutoff and *k*-point mesh were set to 500 eV and 7×7×1, respectively[41]. A vacuum region of 20 Å was used to avoid the periodic interaction. The stress force and energy convergence criteria were chosen as 0.01 eV/Å and $10^{-5}$ eV, respectively. The van der Waals interactions were included in the dissociation process and treated by the semi-empirical DFT-D3 method[42,43]. A supercell of 8×8 graphene unit cells was adopted for the simulations, and ripples were characterized by the ratio $t/D$ of their height $t$ to the corrugation diameter $D$ (inset of *fig. S6a*). The energy barrier for the reaction pathway was calculated using the climbing-image nudged elastic band (CI-NEB) method, in which the total energies of initial, final and several intermediate states during the reaction process were calculated explicitly[44]. The initial state was constructed as follows[22]. First, we created a corrugated graphene supercell with a certain $t/D$ by allowing the atomic structure to relax under biaxial compression. Next, two hydrogen atoms were attached to specified carbon atoms, and the whole system was allowed to relax to its ground state, during which the positions of unoccupied carbon atoms were fixed to keep the $t/D$ value constant. The relaxed carbon structure was then used as the initial configuration and the electron distribution was optimized during the reaction process.

For a given $t/D$, there are many possible corrugated configurations. If we consider high-symmetry configurations, the corrugation center is located either at the top of a carbon atom or between two nearest neighbors or at the hexagon center. In order to minimize the dissociation energy, we relaxed the above three structures of rippled graphene with two adsorbed hydrogen atoms and used them as the initial states before hydrogenation. The initialized graphene ripple could be allowed to relax further before chemical reaction, but we found that this caused little effect on the energy barrier. After trying many different configurations and reaction processes, we found that the dissociation energy reached minimum if two opposed sites in a hexagon were hydrogenated (see the insets of *fig. S6a*). In this figure, we show changes in the total energy during the reaction process for $t/D$ = 7.5%. The dissociation energy barrier is of about 1.1 eV and given by the difference between the initial and highest energy states along the reaction pathway. For comparison, the dissociation energy of $H_2$ in vacuum is considerably higher, of about 4.5 eV[45], which shows that ripples are highly catalytically active.

The dissociation energy depends on where in the unit cell hydrogen atoms are adsorbed. For example, *fig. S6b* shows the adsorption process for the same *t/D* as in *fig. S6a* but with hydrogen atoms attached to the nearest carbon atoms. In this case, the dissociation energy barrier is higher (~2.9 eV). Our results for different ripple curvatures *t/D* are plotted in *fig. S6c*. Clearly, the dissociation energy decreases monotonically with the curvature, and the changes become rather gradual for *t/D* > 4%. Note that the critical curvature, at which ripples become energetically favorable for dissociation of molecular hydrogen (*t/D* ≈ 2.5%), is somewhat smaller than *t/D* ≈ 4%, which was reported in the earlier study[22]. This is because of improvements in the simulation method and optimized atomistic configurations.

Although graphene membranes are known to contain numerous extrinsic (static) ripples[24,25] that have typical *t/D* ≈ 5%, it is instructive to find what kind of intrinsic (dynamic) ripples one can expect due to thermal fluctuations[26]. To this end, we have performed molecular dynamics simulations using Large-scale atomic/molecular massively parallel simulator (LAMMPS)[46] and graphene membranes consisting of 387,200 atoms. Periodic boundary conditions were usually employed to mimic an infinite membrane, but we also performed simulations for finite-size membranes (from ~35 to 100 nm in diameter). The lateral size of ripples (*D*) ranged between a few and 10 nanometers, independently of the membrane size. Their typical configurations at different *T* were obtained after thermalization in 100,000 steps (0.00025 fs per step) and averaging over twenty of such snapshots (*fig. S7a*). *Figs. S7b and S7c* show the areal density for ripples with *t/D* > 4% which are most catalytically active. One can see that thermal fluctuations generate many such ripples that can result in dissociation of $H_2$.

It is not clear whether static or dynamic ripples dominate the adsorption-dissociation process for graphene membranes. One of the issues limiting a contribution from thermally-excited ripples could be their relatively short lifetimes. Our simulations show that the mean half-life of a ripple, during which its *t/D* drops to half, is of the order of femtoseconds. For comparison, permeation of adsorbed hydrogen atoms through graphene involves the timescale *τ*, which can be estimated from their adsorption energy as $E_{ad} = h/τ$ where *h* is the Planck constant. For atomic hydrogen on graphene[47,48], $E_{ad}$ is expected to be ~0.4–1.0 eV, and this yields *τ* of a few femtoseconds. This is of the same order of magnitude as the characteristic lifetime of ripples.

Besides static and dynamic ripples, there are strained areas around the rim of our microcontainers, which in principle could also contribute to the observed hydrogen permeation. However, this scenario is ruled out by the experimental fact that the observed permeability was proportional to the membrane's area rather than its circumference. Furthermore, permeation rates were same for devices with different sagging (varying from ~5 to 40 nm), which led to different strain. The above conclusion is also supported by our density-functional-theory calculations in which the chemisorption process was considered for flat graphene under strain. *Fig. S6d* shows that the dissociation energy remains large (~3 eV) even for strains as high as ~15%. This proves that it is the curvature rather than strain which is important for the chemisorption process.

**Isotope effect.** The proposed mechanism of hydrogen permeation involves several steps: dissociation of $H_2$ on graphene ripples leading to a finite coverage of the surface with hydrogen atoms; flipping of the atoms across the membrane in a proton-like transfer process; and their recombination and desorption as $H_2$. Only the chemisorption and flipping are expected to involve sufficiently high energy barriers as discussed in the main text. To elucidate which of these two key barriers limits the observed permeation, we performed experiments using deuterium. To this end, ten microcontainers were sealed with monolayer graphene and exposed to $D_2$ at 1 bar at room *T*. Δ𝛿 was monitored as a function of time. In stark contrast to our $H_2$ experiments, the devices did not show any discernable changes in Δ𝛿 (*fig. S8*). The absence of $D_2$ permeation was further verified in elevated-*T* tests such as those in Fig. 2a. Several devices were exposed to $D_2$ at 50°C for three days but none of them exhibited any bulging, in contrast to the hydrogen experiments of Fig. 2a.

To understand the isotope effect, we carried out density-functional-theory calculations for chemisorption of $D_2$ on rippled graphene. After including corrections due to zero-point oscillations, we found little difference in the dissociation energies with respect to $H_2$ (*fig. S6*). On the other hand, desorption of $D_2$ from the graphene surface should be slower because of the same quantum corrections. Hence, surface coverage for deuterium atoms should be higher than that for hydrogen. The latter isotope effect is well known for both graphene and graphite[47,48] and implies that, if chemisorption were the rate-limiting process, higher permeation rates would have been expected for $D_2$ rather than $H_2$, contrary to our observations. The latter conclusion indicates again that flipping is the limiting process. Indeed, in this case, one expects an isotope effect analogous to that observed in the transport experiments where deuterons exhibited much lower conductivity through graphene than protons[12,49]. The reduction was attributed to the fact that deuterons have a lower energy in their initial bound state[12], which results in an effectively higher flipping barrier. The magnitude of the isotope effect depends on the binding energy $E_{ad}$, which for the case of hydrogen adatoms on graphene is expected[47,48] to be ~0.4–1 eV, notably larger than $E_{ad} \approx 0.2$ eV in ref. 12. Accordingly, a larger isotope effect is expected in the present experiments, as compared to that in refs. 12 and 49. This supports the scenario in which the flipping limits hydrogen permeation through monolayer graphene.

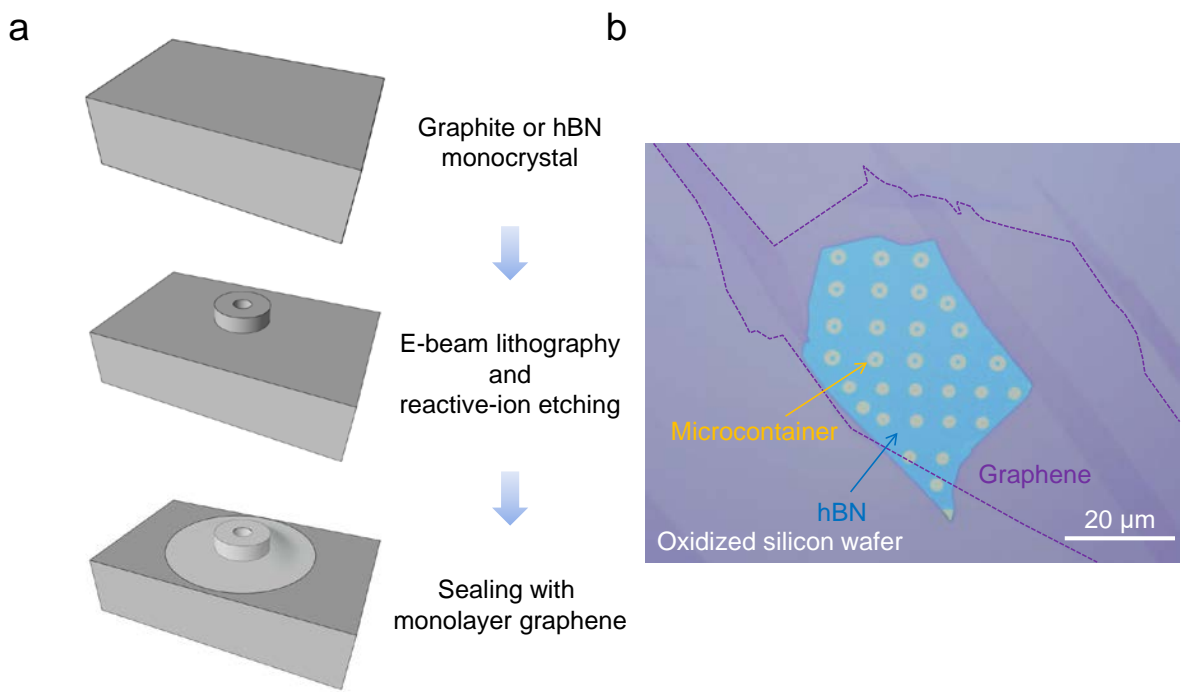

***Figure S1| Device fabrication. a,*** *Graphite or hBN monocrystals are obtained by mechanical exfoliation. Micrometer-size wells are then made by e-beam lithography and ion etching. Monolayer graphene is transferred on top to seal the wells.* ***b,*** *Optical micrograph of a set of hBN microcontainers. The dashed curve indicates the position of monolayer graphene.*

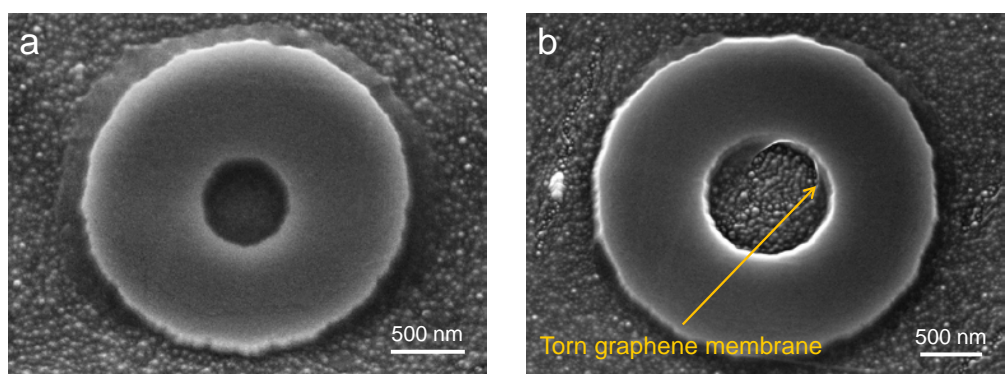

***Figure S2| Electron micrographs of our microcontainers. a,*** *One of them having d = 0.5 μm. Such images were taken only after finishing measurements to avoid electron-beam damage.* ***b,*** *Example of a broken graphene seal: the membrane got damaged after thermal cycling to 80 °C.*

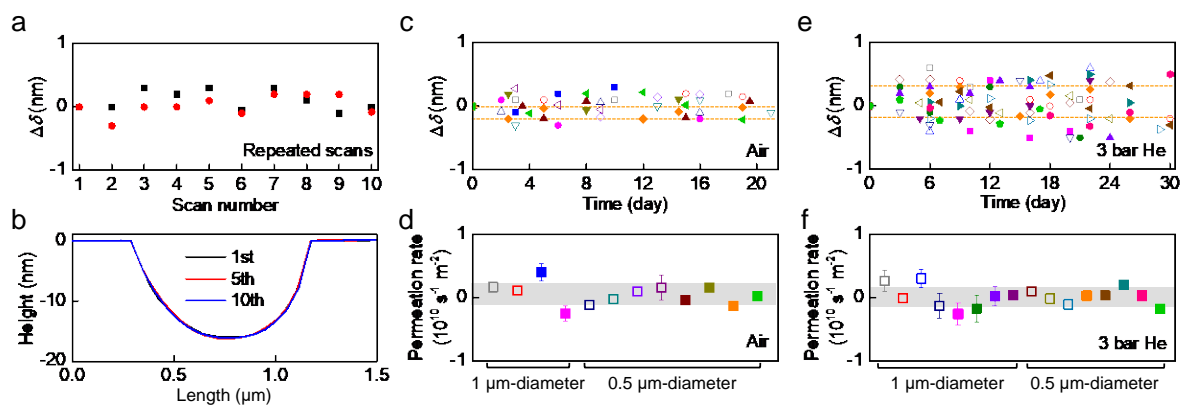

***Figure S3| Stability of graphene membranes in air and helium. a,*** *Changes in $\delta(0)$ measured for two containers with d = 1 μm; AFM scans were taken every hour.* ***b,*** *Representative profiles $\delta(x)$ for (a).* ***c,*** *Long-term variations in $\delta$ for 12 different containers kept in air.* ***d,*** *Permeation rates evaluated from the evolution of $\Delta\delta$ with time in (c).* ***e,*** *$\Delta\delta$ for 16 different devices placed in 3 bar He.* ***f,*** *Permeation rates for the data in (e). In (c) and (e), different symbols denote different microcontainers made from graphite (empty symbols) and hBN (solid). The dashed lines in (c, e) indicate maximum changes detected for representative devices (color coded). In (d) and (f) the color represents the same-color device as in (c) and (e), respectively. Error bars: SD for fitting $\delta$ with a linear time dependence. Grey areas: Overall statistical accuracy obtained using all our devices measured in air and 3 bar He.*

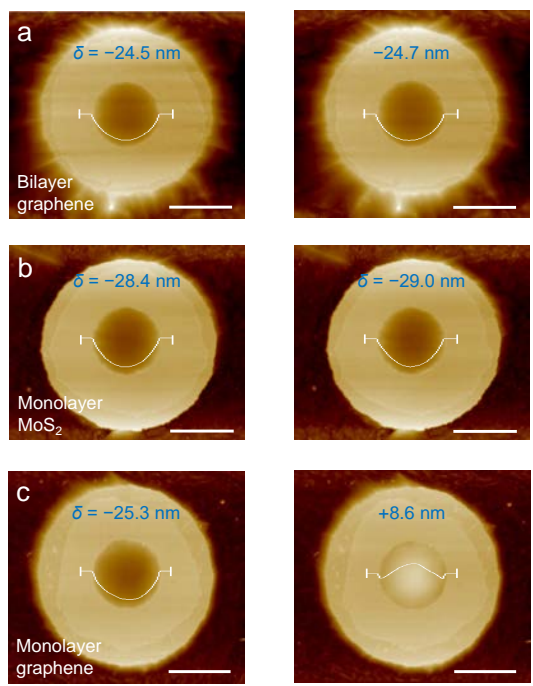

***Figure S4| Bilayer graphene and monolayer MoS$_2$ are impermeable to hydrogen. a,*** *AFM micrographs of the same container sealed with bilayer graphene before (left panel) and after (right) its exposure to 1 bar H$_2$ at 50°C for 3 days. White curves show the profiles along the membrane's diameter. No changes in membrane positions could be detected within our experimental accuracy.* ***b,*** *Same experiment for monolayer MoS$_2$. No changes could be noticed either.* ***c,*** *For comparison, we show the simultaneous experiment for a microcontainer covered with monolayer graphene. The membrane clearly bulged out after the exposure, similar to the case of Fig. 2a of the main text. All scale bars, 1 μm. After the experiment, the bulging membrane in (c) was kept under ambient conditions and found to slowly deflate over months, in agreement with the room-T permeation rates reported in the main text.*

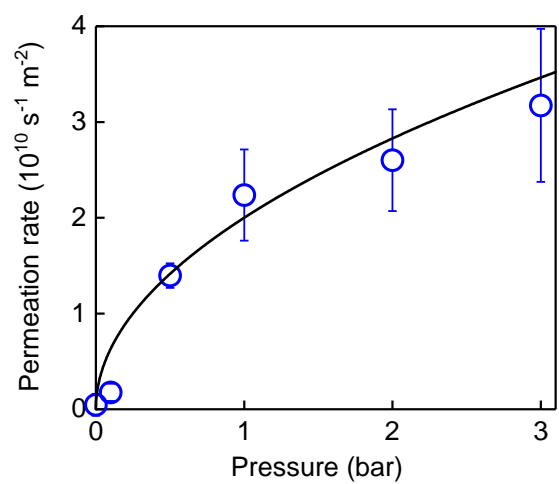

*Figure S5| Pressure dependence of hydrogen permeation.* Symbols: Measurements at room T. Error bars: SD using minimum 10 devices in each case. Solid curve: Best fit to the square-root dependence.

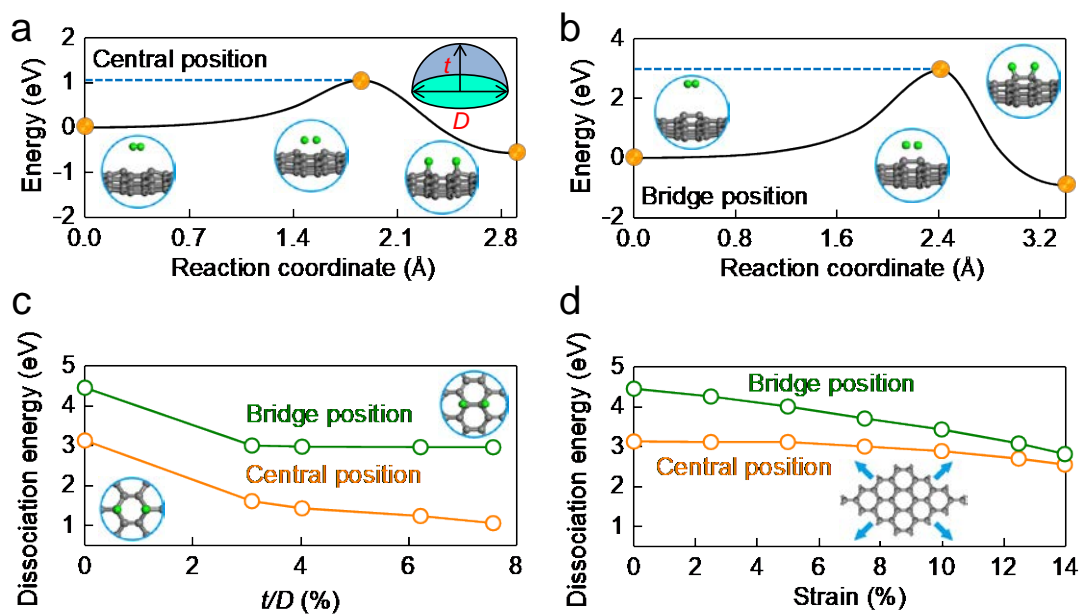

*Figure S6| Dissociation of molecular hydrogen at graphene ripples.* **a** and **b**, Reaction of $H_2$ with graphene for $t/D$ = 7.5% if adatoms are adsorbed in the central and bridge positions, respectively. The insets show atomic configurations of the initial, maximum-energy and final states (marked by the orange dots). **c**, The dissociation barrier as a function of ripples' curvature. Insets: Top-view of the bridge and central positions for hydrogen adatoms. **d**, Dissociation energy as a function of biaxial strain. Inset: Schematic showing the direction of applied strain in our simulations.

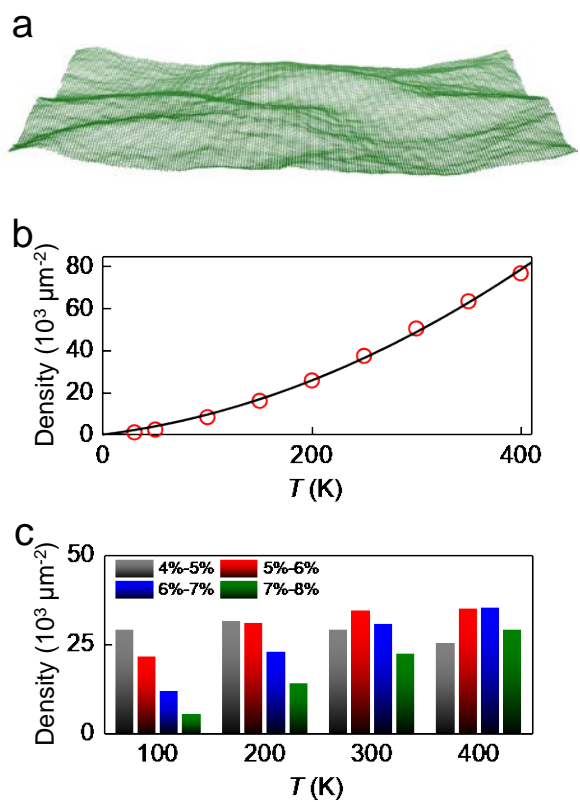

*Figure S7| Intrinsic (dynamic) ripples in graphene at different temperatures. a, Typical snapshot of graphene membrane at 300 K using molecular dynamics simulations. b, Density of ripples with t/D ≥ 7% (most chemically active). Symbols: Calculations for different T. Solid curve: Guide to the eye. c, Statistical distribution of intrinsic ripples with different t/D.*

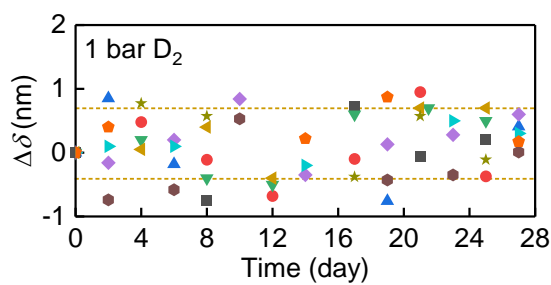

*Figure S8| Isotope effect using deuterium. Time evolution of $\Delta\delta$ for 10 different devices (different colors) exposed to 1 bar of $D_2$ at room T = 295 ±2 K. All the devices are hBN containers sealed with monolayer graphene. The dashed lines indicate maximum changes detected for the device coded with the same color. The random fluctuations are close in amplitude to those shown in Fig. 1e for helium.*